\documentclass[11pt, a4paper]{article}

\usepackage{jheppub}
\usepackage[T1]{fontenc}
\usepackage{graphicx}
\usepackage{amssymb}
\usepackage{subfigure}
\usepackage{hyperref}
\usepackage{mathrsfs}
\usepackage{latexsym}
\usepackage{amsfonts}
\usepackage{amsmath}
\usepackage{esint}
\usepackage{amsxtra}
\usepackage{pifont}

\def \ep1{\epsilon_1}
\def \ep2{\epsilon_2}
\def \m{\mbox}

\def \be{\begin{equation}}
\def \ee{\end{equation}}
\def\beq{\begin{eqnarray}}
\def\eeq{\end{eqnarray}}
\def \ba{\begin{array}}
\def \ea{\end{array}}
\def\nn{\nonumber}

\def \f{\frac}

\def \p{\partial}

\def \sn{\mbox{sn}}
\def \cn{\mbox{cn}}
\def \dn{\mbox{dn}}
\def \ep{\epsilon}
\def \mt{\tilde{m}}

\title{\boldmath N=2 supersymmetric QCD and elliptic potentials}

\author{Wei He}

\affiliation{Instituto de F\'isica Te\'orica, Universidade Estadual Paulista,\\ Barra Funda, 01140-070, S\~{a}o Paulo, SP, Brazil}

\emailAdd{weihephys@gmail.com}

\abstract{We investigate the relation between the four dimensional N=2 SU(2) super Yang-Mills theory with four fundamental flavors and the quantum mechanics model with Treibich-Verdier potential described by the Heun equation in the elliptic form. We study the precise correspondence of quantities in the gauge theory and the quantum mechanics model. An iterative method is used to obtain the asymptotic expansion of the spectrum for the Schr\"{o}dinger operator,
we are able to fix the precise relation between the energy spectrum and the instanton partition function of the gauge theory. We also study asymptotic expansions for the spectrum which correspond to the strong coupling regions of the Seiberg-Witten theory.}

\begin{document}
\maketitle
\flushbottom

\section{Introduction}

Recently we have witnessed some surprising connection between supersymmetric gauge theories, conformal field theories and integrable theories.
Among the many implications of physical and mathematical interests,
we gain some new understanding about the nonperturbative dynamics of quantum gauge theories.
The four dimensional $N=2$ super Yang-Mills theory provides a major playground for these connections since the development of the Seiberg-Witten theory \cite{SW9407,SW9408}.
Its connections to string theory, classical integrable system were among the major concerns for physics interest \cite{integsys1,integsys2, integsys3}.
The instanton counting \cite{instcount} provides an essential quantum field theory explanation for the Seiberg-Witten solution,
and as recently revealed it also provides a quantum generalization for the corresponding integrable system \cite{NS0908}.
The relation of 4D gauge theory/quantum integrable model is part of the program recently outlined by Nekrasov and Shatashvili,
relating the vacuum space of various gauge theories to the Bethe states of some quantum integrable systems \cite{NS0901a, NS0901b}.

In some very simple cases, the integrable system reduces to some simple quantum mechanical problem.
Apply the conventional methods of quantum physics, we can compute the quantum corrections, obtain the spectral property of the quantum model.
Through this kind of quantization we gain a deformed version of the Seiberg-Witten theory,
therefore in this context the deformation in the Nekrasov partition function has a clear physical meaning.

The relation between gauge theories and integrable systems has various interesting implications,
it reveals some structures not explicitly manifested in the original formulation of the two kinds of theories.
However, for the moment our understanding of their relation is based on some sporadic examples we have studied,
there is not a dictionary allowing us to build the precise relation between a given gauge theory and an integrable system, or vice versa.
Therefore a closer study of some particular examples is still worthwhile to better understand how the correspondence works.

In this paper we investigate one example of the gauge theory/integrable model correspondence,
namely the N=2 SU(2) gauge theory with matters and their relation to the spectral theory of some integrable elliptic potentials or their trigonometric limits.
Our goal of the investigation is to find the precise relation between certain quantities on the two sides,
especially to match the infrared dynamics of the gauge theory and the eigenvalue spectrum of the elliptic potential.

The typical example is the SU(2) $N_f=4$ QCD and the associated $BC_1$  Calogero-Inozemtsev (CI) model,
i.e. the quantum mechanical model described by the Heun differential equation \cite{heun, NIST}.
The direct connection between the two theories was not noticed
during the study of classical integrability of Seiberg-Witten
theory \cite{HokerPhong}, recently it appears to be clear partly due
to the relation between gauge theories and CFT which provides
another perspective to the gauge theory/integable model
subject \cite{agt}. The Heun equation in the elliptic form describes
a quantum particle in the Treibich-Verdier(TV) elliptic potential \cite{tv},
which is an interesting topic in its own right. The potential, with
its first appearance in Darboux's work \cite{darboux}, is a notable
object in modern study of dynamical system. It belongs to the so
called integrable finite-gap potential, has close relation to KdV
soliton theory and algebraic-integrability theory \cite{finitegap}.
We hope the study about its relation to SYM theory would enrich the
subject.

The quantum mechanical model, as we call it, is not Hermitian, however it fairly makes sense in the context of algebraic integrable system.
The methods of the Hermitian quantum mechanics apply as well, actually the tools of complex analysis are very useful, as we show in the subsequent sections.
We use ``quantum correction" to refer to the $\ep$-corrections to both the quantum CI model and the gauge theory.

The plan of the paper is the following.

The Section \ref{QCDHeun} is devoted to some necessary background.
We briefly explain how the SU(2) $N_f=4$ gauge theory is related to the Heun
equation through the AGT correspondence, hence we can identify the
relation between the mass parameters of gauge theory and the
coupling strength of the TV potential, given in (\ref{bi}). The
couplings take the form of finite-gap potential.  We explain an
exact WKB method for some linear spectrum problems. For the Heun
equation, a shift of the parameters, manifested as a relation
between the WKB expansion of two functions $\Theta$ and $\Xi$ in
(\ref{theta2xi}), is useful in our discussion.

In the Section \ref{TVspectrum} we first apply the exact WKB
quantization method to obtain the spectrum of the TV quantum model
for the case when the kinetic energy is very large compared to the
potential. The contour integral of the WKB perturbation is carried
out by applying an iterative method, we obtain the perturbative
expansion for the integral for general mass parameters. We then
relate the perturbative spectrum of the TV potential and the low
energy solution of the super QCD theory. The relation between the
energy eigenvalue and the deformed prepotential of gauge theory is
given in (\ref{theta2prepotential}), by matching low order
perturbations on the two sides. A consistent shift argument allows
us to extend the match to higher order perturbations, therefore
predict the full asymptotic eigenvalue for the Heun equation from
the Nekrasov partition function of the super QCD, given in
(\ref{XIallorder}).

In the Section \ref{Dualexpan}  we study another perturbative expansion for the spectrum where the kinetic energy is small compared to the potential,
given in (\ref{delta2adgenericmass}).
It corresponds to the strong coupling expansion of the gauge theory, the procedure of determining expansion point and performing the WKB contour  integral follows the spirit of the Seiberg-Witten theory.

In the Section \ref{Limits} we discuss various limit cases where few other SU(2) gauge theory models are recovered.

In the Appendix \ref{AppHeun} we give the Heun equation in different forms, in accordance with the convention we use,
they are suitable to explain different aspects about its relation to the gauge theory and the quantum mechanical model.
In the Appendix \ref{AppP4} we explain the iterative method to solve the polynomial equation $P_4(z)=0$ in order to carry out the elliptic integral.
In the Appendix \ref{Appstationary} we determine the locations of the stationary points for the elliptic potential, they are in one-to-one correspondence with the singularities in the gauge theory moduli space.

\section{SU(2) $N_f=4$ super QCD and spectrum of elliptic potential \label{QCDHeun}}

\subsection{Schr\"{o}dinger equation from AGT correspondence}

There is not an obvious way to derive the Heun equation from any aspects of the gauge theory,
it appears through a recently discovered relation between N=2 SU(2) gauge theory and Liouville CFT, the Alday-Gaiotto-Tachikawa (AGT) correspondence \cite{agt}.
The AGT states a relation between Nekrasov partition function of $N=2$ gauge theory and the chiral part of the conformal block of Liouville CFT,
both associated to certain punctured Riemann surfaces.  For SU(2) $N_f=4$ supersymmetric QCD,
the relevant instanton partition function for gauge theory in the $\Omega$ background $(\ep_1,\ep_2)$ is related to the 4-point conformal block on the sphere.

In CFT theory, the degenerate operators satisfy constraint conditions of Virasoro generators. Inserting a degenerate operator in the correlator results in certain differential equation \cite{bpz}. Sometimes these differential equations are very useful, for example,
along this line in the minimal models of CFT the 4-point correlator including one degenerate operator can be solved through the hypergeometric equation \cite{bpz, cft}. In the AGT context, we can do the similar thing for Liouville CFT, and the corresponding gauge theory is also affected.
It is argued that the CFT correlator with an additional degenerate operator inserted is related to the N=2 gauge theory with surface operator \cite{loop1},
and the resulting conformal block/Nekrasov partition function in the Nekrasov-Shatashvili (NS) limit is related to the eigenfunction of the corresponding quantum integrable system \cite{at1005}.

In the case of 4-point correlator with an extra degenerate operator constrained by level two Virasoro operators,
the procedure results in a second order partial differential equation for the 5-point conformal block \cite{bpz}.
In the NS limit $\epsilon_1=\ep, \epsilon_2=0$, which is relevant for the quantum integrability, the equation becomes the normal form of the Heun equation \cite{mt1006},
\be (-\ep^2\p_z^2+U(z,m,\ep,\Theta))\Phi(z)=0,\label{equationVep}\ee
where \beq U(z)&=&\f{\mt_1^2-\f{\ep^2}{4}}{z^2}+\f{m_1(m_1-\ep)}{(z-q)^2}+\f{m_0(m_0-\ep)}{(z-1)^2}\nn\\
&\quad&-\f{m_0(m_0-\ep)+m_1(m_1-\ep)-\mt_0^2+\mt_1^2}{z(z-1)}
-\f{(1-q)\Theta}{z(z-q)(z-1)}.\label{potentialep}\eeq
In this equation $z$ is the position of the degenerate operator and it takes complex value, and $\Theta$ is the accessory parameter.
All the parameters are related to gauge theory quantities through the AGT correspondence.
 $m_0, \mt_0, m_1, \mt_1$ are parameters determining the conformal weight of the four primary operators of CFT, they are related to the physical mass of the four flavors $\mu_i$ in gauge theory as\footnote{In the AGT paper \cite{agt} the flavors $\mu_1,\mu_2$ are in the antifundamental representation and $\mu_3,\mu_4$ are in the fundamental representation. In the Nekrasov partition function, the fundamental matter of mass $\mu$ contributes  the same factor as the antifundamental matter with mass $\ep_1+\ep_2-\mu$ contributes. Therefore the parameters in AGT paper \cite{agt}, in the NS limit, are related to ours by $m_0=\ep-m_0^{agt}, \mt_0=-\mt_0^{agt}, m_1=m_1^{agt}, \mt_1=\mt_1^{agt}$. But note that they appear in the equation as $m_0(m_0-\ep)$ and $\mt_0^2-\f{\ep^2}{4}$ unaffected. }
\beq m_0&=&\f{1}{2}(\mu_1+\mu_2),\qquad \mt_0=\f{1}{2}(\mu_1-\mu_2),\nn\\
m_1&=&\f{1}{2}(\mu_3+\mu_4),\qquad \mt_1=\f{1}{2}(\mu_3-\mu_4).\label{massrelation}\eeq
The singularity parameter $q$ is the cross ratio of the position of the four non-degenerate  primary operators in CFT,
identical to the UV coupling of the super QCD theory.

We can further rewrite the equation (\ref{equationVep}) in the
elliptic form, see e.g. \cite{tai1008, kpt2013, NIST}. Define the new coordinate $x$ through \be
z=\f{\wp(x,p)-e_2(p)}{e_1(p)-e_2(p)},\label{z-x}\ee and the new function $W(x)$
by\footnote{The identification of mass parameters to the
$\gamma,\eta,\lambda$ in the equation is not unique, see the
Appendix \ref{AppHeun}. The coefficients for the equation in the
elliptic form depends on the identification. We choose the first
column identification in (\ref{mass2coupling1}).} \be
\Phi(z)=(\wp(x)-e_1)^{-\f{m_0}{\ep}+\f{1}{4}}(\wp(x)-e_2)^{-\f{\mt_1}{\ep}-\f{1}{4}}(\wp(x)-e_3)^{-\f{m_1}{\ep}+\f{1}{4}}W(x),\ee
where $\wp(x)$ is the double periodic Weierstrass elliptic function.
The nome $p=\exp(2\pi i\f{\omega_2}{\omega_1})$,
where $2\omega_1,2\omega_2$ are the periods of $\wp(x)$.
$p$ is related to the instanton counting parameter $q$ in (\ref{potentialep}) by a relation
\be q=\f{\theta_2^4(p)}{\theta_3^4(p)}.\label{qprelation}\ee
Define $\omega_0=0, \omega_3=\omega_1+\omega_2$, we have $e_{1,2,3}=\wp(\omega_{1,2,3})$.
Then we get the following form of the equation, \be
\f{d^2}{dx^2}W+\left(E-\sum_{i=0}^{3}b_i\wp(x+\omega_i)\right)W=0.\label{ellpiticpotential}\ee
It is an eigenvalue problem for the Schr\"{o}dinger operator, the multi-component elliptic potential is the
Treibich-Verdier potential \cite{tv}.

Here we should emphasize a difference between the Lam\'{e}
potential and the TV potential, concerned about the nome of the
elliptic function. For the Lam\'{e} potential which is related to the $N=2^*$ gauge theory, the nome of the
function $\wp(x;\omega_1,\omega_2)$ is identified with the instanton
counting parameter $q$, which is also the modulus of the torus in the M5-brane
construction \cite{g0904}. However, for the TV potential the
nome of the elliptic function is $p$, it is related to the instanton counting
parameter $q$  by the relation (\ref{qprelation}).
In the brane construction $q$ is the cross ratio of the base punctured sphere while $p$ is the modulus of the covering elliptic curve \cite{g0904}.

The Heun equation is equivalent to the quantum $BC_1$  Calogero-Inozemtsev model which is the simplest case of integrable models of Calogero type associated to the $BC_N$ Lie algebra. The coupling strengths are related to the gauge theory mass parameters,
\beq b_0&=&(\f{2\mt_0}{\ep}-\f{1}{2})(\f{2\mt_0}{\ep}+\f{1}{2}),\qquad
b_1=(\f{2m_0}{\ep}-\f{3}{2})(\f{2m_0}{\ep}-\f{1}{2}),\nn\\
b_2&=&(\f{2\mt_1}{\ep}-\f{1}{2})(\f{2\mt_1}{\ep}+\f{1}{2}),\qquad
b_3=(\f{2m_1}{\ep}-\f{3}{2})(\f{2m_1}{\ep}-\f{1}{2}).\label{bi} \eeq
The eigenvalue $E$ is related to the accessory parameter by \beq
E&=&4(e_1-e_3)\f{\Theta}{\ep^2}+f(\f{m_0}{\ep},\f{m_1}{\ep},\f{\mt_0}{\ep},\f{\mt_1}{\ep},e_{1,2,3}),\label{E2theta}\eeq
where the precise form of the function $f$ can be recovered from
(\ref{Q2Theta}), (\ref{H2Q}) in Appendix \ref{AppHeun}, and the
modulus of $e_i$ is $p$. Therefore we also call $\Theta$ as energy
eigenvalue although the normal form of the Heun equation does not
display it as eigenvalue. $\Theta$ is a function of
$m_0^2,m_1^2,\mt_0^2,\mt_1^2, q, \ep$, and the quantum number we
call it $\nu$, the precise relation can be computed by the WKB method
and related to gauge theory partition function.

Note that the map (\ref{z-x}) is not one-to-one, in the complex plane every fundamental region of the $x$-plane is mapped to the whole complex plane of $z$.
The points of $x=\omega_1,\omega_2,\omega_3, 0$, module periods, are mapped to $z=1, 0, q, \infty$, respectively.

Therefore we get the direct relation between the supersymmetric QCD theory and the quantum model of Treibich-Verdier potential, we will analyse the asymptotic spectrum of the potential and its relation to the duality of gauge theory.  Actually, the Heun equation satisfied by the 5-point conformal block with one  degenerate operator was already discovered for a different purpose, in the Weierstrass form \cite{flno0902}.
This equation can be derived directly following the original paper of Belavin, Polyakov and Zamolodchikov \cite{bpz}, and has been used in many papers in different forms, there are few recent study relevant to the Heun equation in the context of relations among conformal field theory, integrable theory and matrix models \cite{tai1008, mmm1011, piatek1102, bmt1102, ferraripiatek2012, kpt2013}.

\subsection{Integrable finite-gap potentials}

The following linear spectrum problem for the potential $u(x)$,
\be -\psi^{''}+u(x)\psi=E\psi,\ee
is an outstanding source of knowledge for subjects from classical functional analysis to quantum physics and integrable theory.
In real analysis, a particular interesting case is when the potential is periodic, $u(x)=u(x+T)$,
then the solution of the equation takes the form of Floquet-Bloch wave function,
\be \psi(x+T)=e^{i\nu T}\psi(x).\ee
For every quantum number $\nu$ the spectrum has a solution $E(\nu)$, however for stable solutions the reverse function $\nu=\nu(E)$ must be real,
i.e. $|\cos\nu T|\leqslant1$, this condition restricts the allowed energy inside some bands in the line of $E$,
with the boundaries determined by the periodic and anti-periodic solution $E(\nu_*)$, with $\nu_*$ determined by
$ e^{i\nu_*T}=\pm1$. Other regions are the forbidden zones where solutions are not stable. As one dimensional model for electrons in a lattice, the gap structure is responsible for the electroconductivity of materials.

Generally, the width of energy gaps decreases as the energy increases, it is possible somewhere in the spectrum the gap disappears.
If moreover the spectrum is bounded from below, then the number of forbidden zones is a finite number, and the potential is called {\em finite-gap potential}.
The finite-gap potentials are studied in the context of algebraic integrability theory where analytical tools of Riemann surface are available \cite{finitegap}.
The Hamiltonian system of the KdV hierarchy generates a family of elliptic finite-gap potentials by the fact that (quasi)periodic solutions of the stationary higher KdV equations are finite-gap potentials \cite{novikov, lax}.
The $g$-gap potentials can be written in term of the Riemann Theta function on the genus $g$ surface \cite{ItsMatveev},
and they are isospectral under the KdV flows.

Interestingly, the elliptic potentials related to 4D SU(2) SYM
theories are among the most studied finite-gap potentials. The
Lam\'{e} potential, acquiring the name from the Lam\'{e}
equation,\be u(x)=g(g+1)\wp(x),\ee is a finite-gap potential with
$g$ forbidden zones in the spectrum when $g$ is an integer, a result
due to E. L. Ince. The Lam\'{e} equation is related to the mass
deformed SU(2) $N=4$ Yang-Mills theory($N=2^*$ theory), we have
analysed their relation in a previous work \cite{wh1108}. The
Treibich-Verdier potential is a generalization of the Lam\'{e}
potential, it is also a finite-gap potential when the coupling
coefficients $b_i$ take the following form, \be
u(x)=\sum_{i=0}^{3}g_i(g_i+1)\wp(x+\omega_i),\ee and $g_i$ are
integers \cite{tv}. When the couplings are identified with the mass
parameters of the supersymmetric QCD, the coupling constants $b_i$
in (\ref{bi}) indeed can be written in the form above, but in
general the corresponding $g_i$ are not integers. Of course, in the
gauge theory there is not a reason to demand $g_i$ to be integers.
If we were satisfied by the relation of gauge theory and the
spectral solution of Schr\"{o}dinger operator then it is not
necessary to impose the integer condition for $g_i$. However, if the
masses take special values that $g_i$ are integers then we can
further relate the $\ep_1$-deformed gauge theory to the classical
elliptic solution of KdV theory. Moreover, the full deformed gauge
theory/quantum CFT with nonzero $\ep_1,\ep_2$ provides a deformation
for the classical KdV solution, as demonstrated for the Lam\'{e}
potential in \cite{wh1401}. In this paper we do not focus on this
point, but similar arguments can be made for the Treibich-Verdier
potential.

There are a few ways to reduce the TV potential to the Lam\'{e} potential.
\begin{itemize}
\item[$\bullet$] The obvious one is by turning any three parameters among $b_0, b_1, b_2, b_3$ to zero.
\item[$\bullet$] We can also turn them to $b_0=b_1=b_2=b_3=b$ and use the
duplication formula for the elliptic function, \be
4\wp(2x)=\wp(x)+\wp(x+\omega_1)+\wp(x+\omega_2)+\wp(x+\omega_3).\label{CIpotential}\ee
\item[$\bullet$] The less obvious case is when $b_i$ take special value,
including: \ding{172} $b_0=b_1\ne0, b_2=b_3=0$ or $b_0=b_1=0, b_2=b_3\ne0$;
\ding{173} $b_0=b_2\ne0, b_1=b_3=0$ or $b_0=b_2=0, b_1=b_3\ne0$. We
can reduce the potential in this class to the Lam\'{e} potential
because we have the following relation by the Landen transformation,
\beq
\wp(x;\omega_1,2\omega_2)&=&\wp(x;2\omega_1,2\omega_2)+\wp(x+\omega_1;2\omega_1,2\omega_2)-\wp(\omega_1;2\omega_1,2\omega_2),\nn\\
\wp(x;2\omega_1,\omega_2)&=&\wp(x;2\omega_1,2\omega_2)+\wp(x+\omega_2;2\omega_1,2\omega_2)-\wp(\omega_2;2\omega_1,2\omega_2),\label{landen}\eeq
we keep the dependence on periods explicit to emphasize the change
of the parameter $\tau=\omega_2/\omega_1$. However, other cases of
$b_0=b_3\ne0,b_1=b_2=0$ or $b_0=b_3=0,b_1=b_2\ne0$ are not allowed. A further
scaling limit reduces the Lam\'{e} equation the Mathieu equation
which is related to the $N=2$ SU(2) pure Yang-Mills theory.
The spectrum of the Mathieu potential $u(x)=\cos x$ contains infinite many gaps.
\end{itemize}

At this point it is also worth to mention that the $BC_1$ CI model/TV potential
is related to the sixth Painlev\'{e} equation. The sixth
Painlev\'{e} equation can be written in the form of a time dependent
Hamiltonian system whose potential is the TV potential \cite{manin}.
Accordingly, the sixth Painlev\'{e} equation is a non-autonomous
version of the classical $BC_1$ CI model, and in fact the relation
to classical Hamiltonian system can be generalised to all other
Painlev\'{e} equations \cite{takasaki0004}.

\subsection{The linear spectrum problem and exact WKB method}

For a linear spectrum problem, a standard method to obtain the perturbative solution is the WKB method when there is an expansion parameter $\ep$.
For a potential with turning points $x_1,x_2$, the quantization condition for the exact wave function is
\be \f{1}{\pi}\int_{x_1}^{x_2}\sum_{n=0}^{\infty}\ep^{n-1}p_n(E,u(x))dx=\nu, \ee
Depending on the boundary condition there may be a requirement about the value of $\nu$, for the periodic potential there is no other requirement other than the gap condition discussed previously. It gives the relation between the eigenvalue and the quantum number $\nu$ and other parameters.
However, for a general potential on the real axis the higher WKB components $p_n$ contain nonintegrable singularities at the turning points.
Some efforts were devoted to overcome this difficult to achieve the exact WKB quantization method \cite{klr1967, bow1977}.
Two useful tricks are introduced:
\begin{itemize}
\item[(1.)] We should extend the integrals to the complex plane, with branch cuts determined by the equation itself.
So the quantization condition becomes a contour integral along the branch cut.
\item[(2.)] In the contour integral, we can trade the higher order integrands $p_n$ by derivatives of less singular functions with respect to parameters (energy, couplings).
The integral and derivative operations are commutative.
Therefore the integration can be performed for the less singular integrands whose singularities become integrable.
\end{itemize}

This kind of exact WKB method works effectively for the trigonometric potential $u(x)=\cos x$, as demonstrated by previous works \cite{MM0910}.
The branch cut plane is the Seiberg-Witten curve in gauge theory, the contour integral of the WKB component $p_0$ is directly related to the contour integral of the Seiberg-Witten form.
The higher order WKB corrections are identified with the Omega background deformation of the gauge theory in the NS limit \cite{NS0908}.

The exact WKB method has been used to investigate the Lam\'{e} potential which is related to $N=2^*$ gauge theory model \cite{wh1108}.
We learned that for a quantum particle moving in an elliptic potential $u(x)=g(g+1)\wp(x)$, at every stationary point of the potential (module periods) there is an asymptotic expansion for the eigenvalue and eigenfunction. In fact the stationary points of the potential are related to the weak and strong coupling regions of the gauge theory where asymptotic expansions for the effective action is available and they are related by electric-magnetic duality.
This paper is a natural continuation of our previous work, here we will analyse the eigenvalue expansion of the TV elliptic potential, expanded as the WKB series.

{\bf Application to the Heun equation}

We start the WKB analysis from the normal form the Heun equation
(\ref{equationVep}), because the integral turns out to be simpler
than other forms of the equation. In some earlier works, e.g.
\cite{mt1006, em1006} the leading order WKB computation is carried
out for simple cases like taking equal mass limit or massless
limit. We extend their analysis to generic mass value, and also
determine other asymptotic expansions which correspond to strong
coupling singularities of gauge theory. However, the potential
$U(z,m,\ep,\Xi)$ involving the Plank constant $\ep$ makes the WKB
analysis more complicated. There is a simple way to take into
account of this effect. We do not need to work on the
Schr\"{o}dinger equation (\ref{equationVep}), instead we start from
the Schr\"{o}dinger equation with the following potential
$V(z,m,\Xi)$, \be
(-\ep^2\p_z^2+V(z,m,\Xi))\Psi(z)=0,\label{equationV0}\ee where \be
V(z)=\f{\mt_1^2}{z^2}+\f{m_1^2}{(z-q)^2}+\f{m_0^2}{(z-1)^2}-\f{m_0^2+m_1^2-\mt_0^2+\mt_1^2}{z(z-1)}
-\f{(1-q)\Xi}{z(z-q)(z-1)}.\label{potential0}\ee The masses
$m_0^2,m_1^2,\mt_0^2,\mt_1^2$ are free parameters, they appear in
the eigenvalue $\Xi$ and the corresponding eigenfunction. We can
shift them by any quantity $\delta m_0^2, \delta
m_1^2,\delta\mt_0^2, \delta\mt_1^2$, then the eigenvalue and
eigenfunction with the new parameters $m_0^2+\delta m_0^2, m_1^2+
\delta m_1^2,\mt_0^2+\delta\mt_0^2,\mt_1^2+\delta\mt_1^2$ still
satisfy the Schr\"{o}dinger equation (\ref{equationV0}). This is of
course true when the shifts take the special values: $\delta
m_0^2=-\ep m_0, \delta m_1^2=-\ep m_1, \delta\mt_0^2=-\ep^2/4,
\delta\mt_1^2=-\ep^2/4$. This fact is already clear in our study of
$N=2^*$ gauge theory/Lam\'{e} potential where the adjoint mass
appears in the shifted form as $m(m-\ep)$  \cite{wh1108}.

Therefore, we can solve the eigenvalue problem of the Schr\"{o}dinger equation (\ref{equationV0}) with potential (\ref{potential0}),
we get the eigenvalue as function of a quantum number $\nu$ and all other parameters, $\Xi(\nu, m_0^2,m_1^2,\mt_0^2,\mt_1^2,q,\ep)$.
Then the eigenvalue $\Theta$ of equation (\ref{equationVep}) with the potential (\ref{potentialep}) takes the same functional form as $\Xi$ but with the mass parameters shifted,
\be \Theta(\nu, m_0^2,m_1^2,\mt_0^2,\mt_1^2,q,\ep)=\Xi\left(\nu, m_0(m_0-\ep),m_1(m_1-\ep),\mt_0^2-\f{\ep^2}{4},\mt_1^2-\f{\ep^2}{4},q,\ep\right) .\label{theta2xi}\ee
This fact not only simplify the computation, we also use this fact to fix the precise relation between the spectrum expansion and the gauge theory partition function,
as we explain later. We emphasize it is the equation (\ref{equationVep}) directly related to gauge theory and CFT.

The WKB form of the wave function is expanded by the Plank constant $\ep$,
\be \Psi(z)=\exp i\int^zdx(\f{p_0(x)}{\ep}+p_1(x)+\ep p_2(x)+\cdots),\ee
The Schr\"{o}dinger equation gives $p_n(z)$ order by order,
\be p_0(z)=i\sqrt{V(z)},\quad  p_1(z)=\f{i}{2}(\ln p_0)^{'},\quad
 p_2(z)=\f{-p_1^2+ip_1^{'}}{2p_0}, \quad \cdots\ee
The potential contains four turning points determined by $p_0(z)=0$, therefore in the complex plane there are two branch cuts and the curve is a torus.

The monodromy of the wave function along the contour $\alpha$ or $\beta$ of the curve can be computed by
\be \nu=\f{1}{2\pi\ep}\oint_{\alpha,\beta} p(z)dz.\label{monod}\ee
$\nu$ can be expanded in accordance with $p(z)$, $\nu=\ep^{-1}\nu_0+\ep\nu_2+\ep^3\nu_4+\cdots$,
and we have $\nu_n=(2\pi)^{-1}\oint p_n(z)dz$.

First, let us work out the leading order monodromy, it is \be
\nu_0=\f{1}{2\pi}\oint_{\alpha,\beta}
p_0(z)dz=\oint_{\alpha,\beta}\f{i}{2\pi}\f{\sqrt{P_4(z)}}{z(z-q)(z-1)}dz,\ee
where $P_4(z)$ is a quartic polynomial of $z$ whose explicit form is
in Appendix \ref{AppP4}. As shown in \cite{mt1006, em1006} it is
simpler to compute the contour integral for $\p_\Xi \nu_0$ first
because $\p_\Xi P_4(z)=-(1-q)z(z-q)(z-1)$, therefore we have \be
\f{\p\nu_0}{\p \Xi}=\f{1-q}{4\pi
i}\oint_{\alpha,\beta}\f{dz}{\sqrt{P_4(z)}}.\label{alphabetainteg}\ee
As $P_4(z)$ is a quartic polynomial, the integral is complete
elliptic integral. The result depends on the four roots of the
equation $P_4(z)=0$. Suppose we have four roots $z_i, i=1,2,3,4$, we
can factorize $P_4(z)$ as $ P_4(z)=\mt_0^2(z-z_1) (z-z_2) (z-z_3)
(z-z_4)$, where the factor $\mt_0^2$ is the coefficient of $z^4$.
Then the integrals are
\beq \int_{z_1}^{z_2}\f{dz}{\sqrt{(z-z_1)(z-z_2)(z-z_3)(z-z_4)}}&=&\f{2i}{\sqrt{(z_1-z_3)(z_2-z_4)}}K(k^2),\nn\\
\int_{z_2}^{z_3}\f{dz}{\sqrt{(z-z_1)(z-z_2)(z-z_3)(z-z_4)}}&=&\f{-2}{\sqrt{(z_1-z_3)(z_2-z_4)}}K(k^{'2}),\label{completelliptic}\eeq
for contour $\alpha$ and $\beta$ respectively. $K(k^2)$ is the complete elliptic integrals of the first kind,
$k$ is the modulus given by the cross ratio of the four roots, and $k^{'}$ is the complementary  modulus.
\be k^2=\f{(z_1-z_2)(z_3-z_4)}{(z_1-z_3)(z_2-z_4)},\qquad k^{'2}=1-k^{2}=\f{(z_2-z_3)(z_1-z_4)}{(z_1-z_3)(z_2-z_4)}.\label{modulus}\ee
The modulus $k^2$ and $q$ are different quantities, $k^2$(or $k^{'2}$) describes the IR coupling of gauge theory while $q$ describes the UV coupling of gauge theory.

In order to get an asymptotic expansion, we need the four roots with
a hierarchical structure that makes either $k^2$ or $k^{'2}$ small.
This happens when two of the roots collide while other two remain at
finite distance, for the case of generic parameters, there are six
possible ways. This can be achieved by turning the parameters in the
polynomial $P_4(z)$, in N=2 gauge theory this is controlled by
moving in the moduli space. In the following, we will discuss the asymptotic
expansions for the spectrum of TV potential, which are related to massless
particles of different U(1) charge in the gauge theory context, therefore we use the gauge
theory terminology referring them as ``electric/magnetic/dyonic"
expansions.

The first order perturbation given by the contour integral $\oint
p_1(z)dz$ does not contribute because the integrand is a total
derivative. Actually, all odd order contour integral are zero
because the integrands $ p_{2i+1}(z)$ are all total derivatives.
Higher order contour integral $\oint p_{2i}(z)dz$ can be generated
from the leading order monodromy, by the action of certain
differential operators with respect to the energy and mass
parameters, as have demonstrated for other
simpler potentials \cite{MM0910, wh1006, wh1108}. However, for the TV potential we have more mass
parameters hence the higher order differential operators are harder to compute, so we restrict to the leading order.
But concerned about the higher order $\ep$-expansion, there is a
consistency condition allowing us to obtain them from the gauge theory side, see the next section. Moreover, another different method using the KdV Hamiltonians is recently developed in our subsequent paper \cite{wh1401} and we briefly discuss in the Conclusion.

\section{Perturbative spectrum of Treibich-Verdier potential\label{TVspectrum}}

\subsection{The expansion for large $\Xi$}

We start from the expansion region that corresponds to the electric expansion of gauge theory where the effective coupling is weak.
The first input from the gauge theory is the identification of parameter $\Theta$, or $\Xi$ after the masses shifted,
with the moduli space which is a large quantity in the electric region, $\Xi\gg m_0^2,m_1^2,\mt_0^2,\mt_1^2$.
Then the contour integral is along the $\alpha$-cycle where roots $z_1\sim0, z_2\sim q$ are close with each other.
The leading order monodromy $\nu_0$ is identified with the v.e.v of the scalar field in the undeformed gauge theory $a_0$.
When gauge theory is deformed, the v.e.v of the adjoint scalar is also deformed as $a=a_0+\ep^2 a_2+\ep^4 a_4+\cdots$.
We identify $\nu$ with $a$ by $\nu=\f{a}{\ep}$, i.e. we have $\nu_n=a_n$.

Let us compute the leading order WKB integral for $\nu_0=a_0$. For large $\Xi$, the equation $P_4(z)=0$ can be iteratively solved order by order in the $\Xi$ expansion if we correctly choose the leading order solution. For details of the method see Appendix \ref{AppP4}. The result is
\beq z_1&=&\f{q\mt_1^2}{(1-q)\Xi}+\f{f_1^2}{\Xi^2}+\f{f_1^3}{\Xi^3}+\cdots,\nn\\
z_2&=&q-\f{qm_1^2}{\Xi}+\f{f_2^2}{\Xi^2}+\f{f_2^3}{\Xi^3}+\cdots,\nn\\
z_3&=&1+\f{m_0^2}{\Xi}+\f{f_3^2}{\Xi^2}+\f{f_3^3}{\Xi^3}+\cdots,\nn\\
z_4&=&\f{(1-q)\Xi}{\mt_0^2}+f_4^0+\f{f_4^1}{\Xi}+\f{f_4^2}{\Xi^2}+\cdots.\label{largeexp}\eeq
The coefficients $f_i^j$ with the subscript denotes roots $i=1,2,3,4$, the superscript denotes the (minus of) power of $\Xi$.
We present the first few $f_i^j$ in Appendix \ref{AppP4}.
The roots have the right hierarchical pattern, $z_1\sim0\ll z_2\sim q\ll z_3\sim1\ll z_4\sim\infty$, then the  modulus $k$ is
\be k^2=q-\f{m_0^2+\mt_0^2+m_1^2+\mt_1^2}{\Xi}q+\mathcal{O}(\Xi^{-2}q)+\mathcal{O}(q^{2}).\ee
It remains small because $q$ is small.
Then the $\p_\Xi \nu_0$ and its $\Xi$ expansion is given by
\beq \f{\p a_0}{\p\Xi}&=&\f{(1-q)}{\pi\mt_0}\f{K(k^2)}{\sqrt{(z_1-z_3)(z_2-z_4)}}\nn\\
&=&\f{1}{2}\Xi^{-\f{1}{2}}(h_0+h_1\Xi^{-1}+h_2\Xi^{-2}+\cdots),\eeq
where $h_i$ are functions of $m_0,m_1,\mt_0,\mt_1, q$.
Integrate $\Xi$, and reverse the series, we get
\be \Xi=\f{a_0^2}{h_0^2}[1+2h_0h_1a_0^{-2}+(\f{2}{3}h_0^3h_2-h_0^2h_1^2)a_0^{-4}
+(\f{2}{5}h_0^5h_3-2h_0^4h_1h_2+2h_0^3h_1^3)a_0^{-6}+\cdots].\label{Theta2a0}\ee
Substituting the expressions of $h_i$, we have
\be \Xi=a_0^2-m_1^2-\mt_1^2+\f{(a_0^2+m_0^2-\mt_0^2)(a_0^2+m_1^2-\mt_1^2)}{2a_0^2}q+\mathcal{O}(q^2).\label{Xileadorder}\ee

The first non-zero quantum correction to $\nu$ is $\nu_2=a_2$. Following the method of earlier work \cite{MM0910}, it would be given by a differential operator acting on the leading order result. We do not proceed here because the coefficients are very lengthy if the masses are of generic value.
Including all quantum effects, the monodromy can be expanded as $\ep^{-1}a=\ep^{-1}a_0+\ep a_2+\ep^3 a_4+\cdots$, $a_n$ are function of $\Xi$.
The inverse gives the expansion of $\Xi$ in the form $\Xi=\Xi_0+\ep^2\Xi_2+\ep^4\Xi_4+\cdots$. with $\Xi_{2n}$ functions of $a$,
instead of $a_0$ \cite{wh1006}. In this form we can relate the function $\Xi(a,m,q,\ep)$ to the Nekrasov partition function of the gauge theory, through the Matone's relation \cite{matone1, matone2}. In the following section, we will show how to match the WKB expansion with the gauge theory result without performing higher order WKB computation.

\subsection{Match with the instanton partition function}

According to the gauge theory/integrable model relation, the moduli parameter of gauge theory is proportional to the energy eigenvalue of integrable model,
the v.e.v of adjoint scalar is identified with the momentum of quasi-particles.
In order to establish the precise relation between the two models, we need to compute the expansions on the two sides, at least for the first few orders.
For N=2 gauge theory, the leading order solution can be computed from the Seiberg-Witten curve,
but this mechanism is unable to get information about quantum corrections.

There is a particular region in the moduli space where the gauge
theory is formulated by a Lagrangian and the coupling is weak,
therefore the theory is in good control by QFT method. The
exponentially suppressed instanton contribution is given by a
counting algorithm \cite{instcount}, incorporating the
$\ep$-deformation by the $\Omega$ background, the $\ep$-corrected
effective action obtained is directly related to the quantum
spectrum of the model. The effective action contains the
perturbative part and the instanton part \cite{instcount, no0306}.
The perturbative part of the SU(2) theory with four flavors is $
Z^{pert}=\exp-\f{1}{\ep_1\ep_2}(a^2\ln q+\cdots)$, where we omit
terms that do not depend on $q$, because only the first term play a
role in the identification (\ref{theta2prepotential}). The instanton
contribution $Z^{inst}=1+Z_1q+Z_2q^2+\cdots$ can be computed from
the Nekrasov partition function \cite{instcount}. The deformed
prepotential is obtained from the instanton partition function by
$\mathcal{F}=-\ep_1\ep_2\ln(Z^{pert}Z^{inst})$, we take the NS limit
$\ep_1=\ep, \ep_2=0$, then we can expand the deformed prepotential
as $\mathcal{F}=\mathcal{F}_{(0)}+\ep\mathcal{F}_{(1)}+\cdots$ where
at each order $\mathcal{F}_{(n)}$ contains the perturbative part and
the instanton part. As emphasized in AGT paper \cite{agt}, the
Nekrasov partition function actually computes the instanton action
of U(2) gauge theory, therefore an U(1) factor need to be subtracted
to get result of SU(2) gauge theory. We use a form of the instanton
computing formula convenient for program treatment \cite{FP0208,
bfmt}.

It is the function $\Theta$ associated to the Heun equation
(\ref{equationVep}) directly related to the gauge theory, therefore
we need to shift the masses parameters of the function $\Xi$ for the
equation (\ref{equationV0}) obtained in the previous section to get
$\Theta$, then relate it to gauge theory. The relation between the
function $\Theta$ and the deformed prepotential of U(2) $N_f=4$
supersymmetric QCD in the limit $\ep_1=\ep, \ep_2\to0$ can be fixed,
\be \Theta(a,
m_0^2,m_1^2,\mt_0^2,\mt_1^2,q,\ep)+m_1(m_1-\ep)+\mt_1^2+\f{2q}{1-q}(m_0-\ep)(m_1-\ep)=q\f{\p}{\p
q}\mathcal{F}(a,\mu_i,q,\ep), \label{theta2prepotential}\ee where on
the left hand side the mass parameters are $m_0,m_1,\mt_0,\mt_1$, on
the right hand side the mass parameters are
$\mu_1,\mu_2,\mu_3,\mu_4$, they are related through
(\ref{massrelation}). This relation, often not written in the form
above, has been checked in equal mass case \cite{mt1006}, used in
discussion of relation to classical conformal block \cite{tai1008,
piatek1102, ferraripiatek2012}. In this paper we carry out a direct
spectral analysis for $\Theta$ when masses take generic value, find
its three asymptotic expansions in the parameter space (relation
(\ref{theta2prepotential}) is one of them), and check these facts
against the gauge theory.

Let us explain how it is determined from the process of obtaining
the equation (\ref{equationVep}) from $\ep_2\to0$ limit of the BPZ
\cite{bpz} equation of Liouville CFT, see e.g. \cite{flno0902,
loop1, mt1006, tai1008, piatek1102, ferraripiatek2012, kpt2013}. The
null decoupling equation is a partial differential equation, the
chiral part of the integrand of the 5-point correlation function is
interpreted as the wave function. In the classical limit
$\ep_2\to0$, the effect of degenerate operator is in subleading terms of the wave function because it is a light
operator, and the equation becomes an ordinary differential
equation. Then $\Theta$ is the accessory parameter of the Fushian
equation, given by the classical limit of chiral integrand of
4-point function of heavy operators \cite{Zamolod9506}. The 4-point
conformal block is further related to the Nekrasov instanton action
by the AGT relation \cite{agt}. Taking account all these relations,
we can fix the relation (\ref{theta2prepotential}). In particular,
the $-a^2+m_1(m_1-\ep)+\mt_1^2$ part comes from
$\Delta_{int}-\Delta_3-\Delta_4$, as explained in \cite{mt1006}, the
term $\f{2q}{1-q}(m_0-\ep)(m_1-\ep)$ comes from the U(1) factor in
AGT relation because in (\ref{theta2prepotential}) we use deformed
prepotential of U(2) gauge theory (already set $a_1=-a_2=a$). We
have confirmed this relation to higher order $\ep$ expansion by a
different method of computing $\Theta$ from the KdV Hamiltonians
\cite{wh1401}.

From this relation we can derive the instanton part of the prepotential of the gauge theory $\mathcal{F}_{(0)}^{inst}$ and $\mathcal{F}_{(1)}^{inst}$ from the leading order result of $\Xi$ in (\ref{Xileadorder}),
and the relation (\ref{theta2xi}) and (\ref{theta2prepotential}).
The first few order results, now written in terms of the physical mass $\mu_i$ used in gauge theory, are
\beq \mathcal{F}_{(0)}^{inst}&=&\f{1}{2a^2}(a^4+a^2\sum\limits_{i<j}\mu_i\mu_j+\prod\limits_{i}\mu_i)q\nn\\
&&+\f{1}{64a^6}[13a^8+a^6(\sum\limits_i\mu_i^2+16\sum\limits_{i<j}\mu_i\mu_j)+a^4(\sum\limits_{i<j}\mu_i^2\mu_j^2+16\prod\limits_{i}\mu_i)\nn\\
&&-3a^2\sum\limits_{i<j<k}\mu_i^2\mu_j^2\mu_k^2+5\prod\limits_{i}\mu_i^2]q^2+\mathcal{O}(q^3),\label{elexp0}\eeq
and
\beq \mathcal{F}_{(1)}^{inst}&=&-\f{1}{4a^2}(5a^2\sum\limits_i\mu_i+\sum\limits_{i<j<k}\mu_i\mu_j\mu_k)q\nn\\
&&-\f{1}{64a^6}[41a^6\sum\limits_i\mu_i+a^4(\sum\limits_{i<j}\mu_i^2\mu_j+\sum\limits_{i<j}\mu_i\mu_j^2+8\sum\limits_{i<j<k}\mu_i\mu_j\mu_k)\nn\\
&&-3a^2\sum\limits_{i<j<k}(\mu_i^2\mu_j^2\mu_k+\mu_i^2\mu_j\mu_k^2+\mu_i\mu_j^2\mu_k^2)+5\prod\limits_{l}\mu_l\sum\limits_{i<j<k}\mu_i\mu_j\mu_k]q^2\nn\\
&&+\mathcal{O}(q^3),\label{elexp1}\eeq where the indices $i,j,k,l\in
\lbrace 1,2,3,4\rbrace$. The $\mathcal{F}_{(0)}$ is the
Seiberg-Witten solution. They precisely agree with the results of
Nekrasov instanton partition function for U(2) $N_f=4$ theory with
generic masses, hence confirm the relation
(\ref{theta2prepotential}) for low order of $\ep$-expansion.

In order to validate the relation (\ref{theta2prepotential}), we should work out few higher order WKB analysis of the Heun equation.
In principle this can be done. However, even without the higher order WKB results there is a nontrivial consistency condition which allows us to proceed further.
The functions $\Xi$ and $\Theta$, expanded as series of $\ep$, have different forms.
Because in the equation (\ref{equationV0}) the potential $V(z)$ does not contain $\ep$, the Schr\"{o}dinger operator is invariant under the change $\ep\to-\ep$,
therefore the spectrum function $\Xi$ contains only even order of $\ep$.
Indeed, for a potential independent of $\ep$ the contour integrals for odd order WKB component $p_{2n+1}$ always vanish. Therefore we have
\be \Xi=\sum_{n}\ep^{2n}\xi_{2n}(a,m_0^2,m_1^2,\mt_0^2,\mt_1^2,q),\qquad n\geqslant0.\ee
But in the equation (\ref{equationVep}) the potential $U(z,\ep)$ involves $\ep$, the integrals of $p_{2n+1}$ are nonzero, the function $\Theta$ should contain all order of $\ep$,
\be \Theta=\sum_{n}\ep^n\theta_n(a,m_0^2,m_1^2,\mt_0^2,\mt_1^2,q),\qquad n\geqslant0.\nn\\\ee

As discussed previously, if we shift the arguments of masses in the
function $\Theta$ in a proper way, then formally we can write it as
a series with only even power of $\ep$, \be
\Theta=\sum_{n}\ep^{2n}\widetilde{\theta}_{2n}\left(a,m_0(m_0-\ep),m_1(m_1-\ep),\mt_0^2-\f{\ep^2}{4},\mt_1^2-\f{\ep^2}{4},q\right),\qquad
n\geqslant0.\label{thetashifted}\ee now with
$\widetilde{\theta}_{2n+1}$ vanish, and the new functions
$\widetilde{\theta}_{2n}$ take the same functional form as
$\xi_{2n}$, merely with arguments shifted. Notice that if we suppose
the relation (\ref{theta2prepotential}) correct, then the function
$\Theta$ can be derived from the deformed prepotential, indeed it
contains terms of all order of $\ep$. It is a nontrivial requirement
that we can rearrange the expansion in such a way where all odd
order $\ep$ terms vanish by shifting the mass parameters.
Especially, from this formula it is clear $\theta_1$ entirely comes
from $\theta_0$ by the shift of $m_0^2$ and $m_1^2$, therefore
$\theta_1=(-m_0\f{\p}{\p m_0^2}-m_1\f{\p}{\p
m_1^2})\theta_0=-\f{1}{2}(\f{\p}{\p m_0}+\f{\p}{\p m_1})\theta_0$.
This is consistent with the $\ep$-order result in \cite{mt1006} by
explicitly examining the first two order WKB perturbation for the
equation (\ref{equationVep}).

The rearrangement indeed works. As the instanton counting can be easily computed in a programmed way,
therefore we can use the relation (\ref{theta2prepotential}) to predict the higher WKB expansion for the function $\Theta$.
Then we shift the masses as explained and get the expansion (\ref{thetashifted}),
from the functional form of $\widetilde{\theta}_{2n}$ we obtain the function $\xi_{2n}$.
For example, the first few order of $\ep$-expansion for $\Xi$ are
\beq \Xi=&&a^2-m_1^2-\mt_1^2+\f{(a^2+m_0^2-\mt_0^2)(a^2+m_1^2-\mt_1^2)}{2a^2}q+\mathcal{O}(q^2)\nn\\
&&+\ep^2\left(-\f{1}{4}+\f{-a^4+(m_0^2-\mt_0^2)(m_1^2-\mt_1^2)}{8a^4}q+\mathcal{O}(q^2)\right)\nn\\
&&+\ep^4\left(\f{(m_0^2-\mt_0^2)(m_1^2-\mt_1^2)}{32a^6}q+\mathcal{O}(q^2)\right)+\mathcal{O}(\ep^6).\label{XIallorder}\eeq
However this does not mean we can shift the masses in the same way to make the odd order $\mathcal{F}_{(2n+1)}$ vanish,
because in the relation (\ref{theta2prepotential}) the term $(m_0-\ep)(m_1-\ep)$ is not in the shifted form, it plays a special role in the rearrangement.

\section{Perturbative spectrum for small $\Delta$\label{Dualexpan}}

According to the duality of the gauge theory, there are other asymptotic expansions given by contour integrals along the dual cycles,
such as $\beta$ and $\alpha+\beta$. Stated in terms of the monodromy of the wave function in (\ref{monod}),
the contour integral along the $\beta$ cycle gives the magnetic dual description.
The dual of $a$ is denoted as $a_D$, deformed as $a_D=a_{D0}+\ep^2 a_{D2}+\ep^4 a_{D4}+\cdots$, now we can identify $\nu_n=a_{Dn}$.
As we actually only work out the leading order, we omit the subscript.
In order to determine the location for other expansions in the strong coupling regions in the moduli space,
we need to find finite value solutions for the six degree equation $\mathcal{D}(P_4)(\Xi)=0$.
In the case of generic mass, the solutions are complicated, however,
asymptotic solutions can be found by the iterative method as explained in Appendix \ref{AppP4}.

{\bf Magnetic expansion}

One of the solutions takes the form which corresponds to the dual magnetic description of the gauge theory dynamics,
\be \Xi_{mag}=-(m_1^2+\mt_1^2)+g_{n}q^{n/2},\qquad n=1,2,3,4,\cdots.\ee
The first few $g_n$ are
\beq g_1&=&2((m_0^2-\mt_0^2)(m_1^2-\mt_1^2))^{1/2},\nn\\
g_2&=&-\f{m_1^4\mt_0^2+m_0^4\mt_1^2-2m_0^2\mt_0^2\mt_1^2-2m_1^2\mt_0^2\mt_1^2+\mt_0^4\mt_1^2+\mt_0^2\mt_1^4}{(m_0^2-\mt_0^2)(m_1^2-\mt_1^2)},\nn\\
\cdots
\eeq
In the mass decoupling limit only the term $\mathcal{O}(q^{1/2})$ survives, it is symmetric w.r.t the masses,
therefore the shifted coordinate $\widetilde{\Xi}_{mag}$, defined in Appendix \ref{AppP4}, is finite under all steps of the successive decoupling limits.

Then we set $\Xi=\Xi_{mag}+\Delta$, with $\Delta$ a small quantity compare to $q^{1/2}(m_0^2-\mt_0^2)^{1/2}(m_1^2-\mt_1^2)^{1/2}=q^{1/2}(\mu_1\mu_2\mu_3\mu_4)^{1/2}$.
Substitute $\Xi=\Xi_{mag}+\Delta$ into $P_4(z)$, the polynomial now involves the small quantity $\Delta$,
then we can continue to iteratively solve the equation $P_4(z)=0$.
The first few order of $z_i, i=1,2,3,4$,  are presented in Appendix \ref{AppP4}.
The roots indeed give us the small complementary modulus as \be
k^{'2}=\f{2\Delta^{1/2}}{(m_0^2-\mt_0^2)^{1/4}(m_1^2-\mt_1^2)^{1/4}q^{1/4}}-\f{2\Delta}{(m_0^2-\mt_0^2)^{1/2}(m_1^2-\mt_1^2)^{1/2}q^{1/2}}+\cdots\ll
1.\ee In $k^{'2}$ we only present the leading order terms that
survive in the full mass decoupling limit which scale as
$(q\mu_1\mu_2\mu_3\mu_4)^{-n/4}$.

Substitute these data to the integral (\ref{alphabetainteg}), we finally get the reverse expansion which would be compact if written in terms of physical masses, it begins as
\beq \Delta&=&\Bigg(-2(\mu_1\mu_2\mu_3\mu_4)^{1/4}q^{1/4}+\f{\sum_{i<j<k}\mu_i^2\mu_j^2\mu_k^2}{2(\mu_1\mu_2\mu_3\mu_4)^{5/4}}q^{3/4}\nn\\
&\quad&+\f{9\sum_{i<j<k}\mu_i^4\mu_j^4\mu_k^4-10(\mu_1\mu_2\mu_3\mu_4)^2\sum_{i<j}\mu_i^2\mu_j^2-40(\mu_1\mu_2\mu_3\mu_4)^3}{32(\mu_1\mu_2\mu_3\mu_4)^{11/4}}q^{5/4}+\mathcal{O}(q^{7/4})\Bigg)\hat{a}_D\nn\\
&\quad&+\Bigg(\f{1}{8}-\f{3\sum_{i<j<k}\mu_i^2\mu_j^2\mu_k^2}{16(\mu_1\mu_2\mu_3\mu_4)^{3/2}}q^{1/2}\nn\\
&\quad&-\f{39\sum_{i<j<k}\mu_i^4\mu_j^4\mu_k^4-30(\mu_1\mu_2\mu_3\mu_4)^2\sum_{i<j}\mu_i^2\mu_j^2-8(\mu_1\mu_2\mu_3\mu_4)^3}{128(\mu_1\mu_2\mu_3\mu_4)^3}q+\mathcal{O}(q^{3/2})\Bigg)\hat{a}_D^2\nn\\
&\quad&+\mathcal{O}(\hat{a}_D^3),\label{delta2adgenericmass}
\eeq
where we have defined $\hat{a}_D=ia_D$, as in \cite{wh1006}. The masses appear in a symmetric way as expected. From the leading order expansion obtained above,
which is the first order in $\Xi=\Xi_{0}+\ep^2\Xi_{2}+\cdots$ if we recover the abbreviated subscript,
we can derive the first order quantum correction to $\Theta=\Theta_0+\ep\Theta_1+\cdots$ which comes from the shift of $m_0^2, m_1^2$,
\be \Theta_0=\Xi_0,\qquad \Theta_1=-\f{1}{2}(\f{\p}{\p m_0}+\f{\p}{\p m_1})\Theta_0=-\sum_{i=1}^{4}\f{\p}{\p\mu_i}\Theta_0,\ee
where $\Theta_0=\Theta_0(\mu_i,\hat{a}_D,q)$ for the dual expansion.

{\bf Dyonic expansion}

The dyonic expansion comes from a rotation of the phase of $q$ by $2\pi$.
From the relation $q=\exp(2\pi i\tau)$ with $\tau=\f{4\pi i}{g_{uv}^2}+\f{\theta}{2\pi}$,
the rotation $q\to e^{2\pi i}q$ induces the shift of the theta angle by $2\pi$,
this would shift the electric charge of a magnetic particle in the low energy gauge theory according to the Witten effect \cite{w1979}, resulting a dyon of charge (1,1).

\section{Various limit cases\label{Limits}}

We provide few limit cases, where we can recover few other gauge theory models and confirm their relations.

{\bf (I). Equal masses limit}

There are few cases the polynomial $P_4(z)$ degenerates to lower order, it is easy to solve $P_4(z)=0$ and the results can be presented in a more compact form. We can take all masses equal $\mu_i=m$, then we have $\mt_0=0, \mt_1=0$.
It is also simple to study the case for $\mu_1=\mu_2, \mu_3\ne\mu_4$ where we have $\mt_0=0$, and the case for $\mu_1\ne\mu_2, \mu_3=\mu_4$ where we have $\mt_1=0$.

{\bf (II). Massless limit}

We can turn some flavors massless, while keep other flavors massive.
This is allowed in the electric expansion, but not allowed in the
dual expansions because this violates our assumption $\Delta\ll q^{1/2}(\mu_1\mu_2\mu_3\mu_4)^{1/2}$. A particular interesting case is the full massless
limit, $m_0=m_1=\mt_0=\mt_1=0$, where the gauge theory becomes
conformal. Let us look at the Seiberg-Witten solution, \be
a_0=-\f{\sqrt{(1-q)\Xi}}{2\pi}\oint_\alpha\f{dz}{\sqrt{z(z-q)(z-1)}}=-\f{2K(q)}{\pi}\sqrt{(1-q)\Xi}.\ee
From $\Xi=q\f{\p\mathcal{F}}{\p q}$ we get the prepotential,
therefore the instanton corrected effective coupling is \be 4\pi
i\tau_{ir}=\f{\p^2\mathcal{F}}{\p
a_0^2}=\f{\pi^2}{2}\int\f{dq}{q(1-q)K^2(q)}=-2\pi\f{K(1-q)}{K(q)}.\ee
If we set $q=\exp(2\pi i\tau_{uv}), p=\exp(4\pi i\tau_{ir})$, then
for weak coupling $|q|\ll1$ we have the relation \be 2\pi
i\tau_{ir}=2\pi
i\tau_{uv}-4\ln2+\f{1}{2}q+\f{13}{64}q^2+\f{23}{192}q^3+\f{2701}{32768}q^4+\cdots,\label{tauiruv}\ee
or the inverse relation \be q=\f{\theta_2^4(p)}{\theta_3^4(p)},\ee
which is exactly the relation obtained earlier in another context \cite{gkmw07}, discussed later by e.g. \cite{agt, em1006}.
It is the same relation of  (\ref{qprelation}). The duality of the
massless gauge theory is encoded in the modular transformation of the Theta functions.

{\bf (III). Mass decoupling limit}

In the infinite mass limit, by turning the UV coupling properly, the resulting theory is a gauge theory with less flavors.
In both electric and magnetic expansions,
in the final results the mass parameters always appear in a symmetric way. This feature makes the mass decoupling procedure straightforward, we can decouple any one, any number, of the four flavors.

For example, if we keep $\mu_{1,2,3}$ finite and turn
$\mu_4\to\infty, q\to 0$ while make $q\mu_4=\Lambda_3$ finite, we get the N=2 gauge theory with $N_f=3$.
By turning $\mu_3,\mu_4\to\infty, q\to 0$ with $q\mu_3\mu_4=\Lambda_2^2$ we get the N=2 gauge theory with $N_f=2$.
And  $\mu_2, \mu_3,\mu_4\to\infty, q\to 0$ with $q\mu_2\mu_3\mu_4=\Lambda_1^3$ gives the N=2 gauge theory with $N_f=1$.
Finally we get the N=2 pure Yang-Mills theory by decoupling all flavors as $\mu_{1,2,3,4}\to\infty, q\to 0$ with $q\mu_1\mu_2\mu_3\mu_4=\Lambda^4$ .
We can compare the decoupling results of formula (\ref{elexp0}) to some previous results \cite{hkp96}.

We can also do the decoupling limit for the magnetic(and dyonic) expansion. For example, we first set all masses equal $\mu_i=m$,  then turn $m\to\infty, q\to 0$ and keep $q^{1/4}m=\Lambda$ finite. In the formula (\ref{delta2adgenericmass}), only the following terms survive in the limit,
\be \Delta_{N_f=0}=-2\hat{a}_D\Lambda+\f{1}{2^3}\hat{a}_D^2+\f{1}{2^{7}}\f{\hat{a}_D^3}{\Lambda}+\f{5}{2^{12}}\f{\hat{a}_D^4}{\Lambda^2}
+\f{33}{2^{17}}\f{\hat{a}_D^5}{\Lambda^3}+\f{63}{2^{20}}\f{\hat{a}_D^6}{\Lambda^4}+\cdots.\ee
In agreement with the result of pure gauge theory \cite{wh1006}, if we scale $\hat{a}_D\to 2\hat{a}_D$ because in our parameterization the singularities in the moduli space are at $\pm2\Lambda^2$.

{\bf (IV). Limits related to $N=2^*$ SYM}

According to the AGT correspondence, the partition function of the N=2
$N_f=4$ QCD is related to the 4-point conformal block on the sphere,
and the partition function of the $N=2^*$ SYM is related to the
1-point conformal block on the torus. It turns out that there is a
surprising relation between the two pairs. We will show that, from
the perspective of relation between gauge theory/CFT/integrable
potential,  these facts are in consistent with that the TV potential
reduces to the Lam\'{e} potential in particular limits.

The relation on the CFT side is reflected through the fact that  the 1-point
correlation for a generic primary operator on the torus is related
to 4-point correlator on the sphere with a special choice of
conformal weight for the primary operators  \cite{flno0902}. With a proper
identification of parameters as in (\ref{bi}), in the NS limit, this
choice makes $m_0=m_1=\mt_1=\f{\ep}{4}$ and $\mt_0$ remains free,
therefore we have \be
b_0=(\f{2\mt_0}{\ep}-\f{1}{2})(\f{2\mt_0}{\ep}+\f{1}{2}),\qquad
b_1=b_2=b_3=0.\ee The TV potential in this limit becomes the
Lam\'{e} potential with a single $\wp(x)$ function. It implies a
partial massless limit for the corresponding  gauge theory,
$\mu_1=\mt_0+\f{\ep}{4},\mu_2=-\mt_0+\f{\ep}{4},\mu_3=\f{\ep}{2},\mu_4=0$.

Meanwhile, a relation between the Nekrasov partition functions of the $N_f=4$ gauge theory and the $N=2^*$ gauge theory is also recently found \cite{poghossian0909}.
The choice of the flavor masses is
\be \mu_1=\f{1}{2}M,\quad \mu_2=\f{1}{2}(M+\ep_1),\quad \mu_3=\f{1}{2}(M+\ep_2),\quad \mu_4=\f{1}{2}(M+\ep_1+\ep_2).\ee
 In the NS limit, it implies the parameters for the elliptic potential are
 \be b_0=b_2=0,\qquad b_1=b_3=\f{M}{\ep}(\f{M}{\ep}-1).\ee
The TV potential in this limit is actually also the Lam\'{e}
potential, because we have the relation (\ref{landen}) for the elliptic function.

\section{Conclusion}

In this paper we investigate the relation between the SU(2) super QCD models and the quantum mechanics models of some elliptic potentials.
We carry out a detailed study of SU(2) gauge theory with four fundamental flavors and the associated spectrum of the Treibich-Verdier potential.
This relation implies a few other cases of gauge theory/elliptic potential correspondence,
by taking various limits on both sides.
Our conclusion of the analysis is that the Coulomb branch low energy dynamics of these SU(2) super QCD theories are equivalent to the spectral problem of elliptic potentials.

We compare various aspects on both sides. For example, the Treibich-Verdier potential has six stationary points, they
correspond to six singularities in the moduli of the gauge theory.
We analyse the asymptotic expansions of the eigenvalue of the Schr\"{o}dinger operator at these singularities, we can match one of the expansions for very large eigenvalue with the instanton action of gauge theory. An iterative method is used to factorize the polynomial $P_4(z)$ in
a proper way to find the asymptotic spectrum expansions for the elliptic potential, applicable to all stationary/singularity points.
This method is practically useful to obtain the dual expansion of gauge theory prepotential.
We can compare various limit cases of our computation to previous literatures.
The study supports the fact that the new parametrisation of the Seiberg-Witten curve, manifested as the potential $V(z)$ in (\ref{potential0}),
indeed captures all the ingredients of the low energy gauge theory as originally formulated \cite{SW9408}.

There are obvious questions related to the problem we have studied.
Can we generalise the gauge theory/elliptic potential relation to general quiver gauge theories with more SU(2) groups?
For general N=2 gauge theory with SU($N_c$) gauge group and $N_f=2N_c$ flavor,
is there a corresponding integrable model with the potential of elliptic form? Can we use this gauge theory/elliptic potential relation to find more finite-gap potentials?

Lastly, concerned about the relation to KdV theory, we recently
indeed made further study where we give another method to derive
deformed gauge theory prepotential in the NS limit from classical
KdV Hamiltonians, and the deformed prepotential for generic value of
$\ep_1,\ep_2$ is interpreted as a ``quantum" deformation of the KdV
Hamiltonians \cite{wh1401}. All the method and arguments can be
directly applied to the TV potential. Especially the integration on
elliptic function in (\ref{ellpiticpotential}) naturally gives the
spectrum/prepotential in terms of quasimodular functions of $p$,
using (\ref{qprelation}) we precisely recover the results of this
paper including (\ref{elexp0}), (\ref{elexp1}). Working with the
normal form of the Heun equation, the higher order WKB analysis for the TV
potential would be technically difficult to obtain, hence we cease
to work out them in this paper. The approach from KdV theory avoids
this problem and higher order $\epsilon$ expression are relatively
easier to derive, however the method can only applies to the
asymptotic electric region \cite{wh1401}. Our conclusion is that the
method developed here and the method from KdV theory are
complementary, their consistency is a encouraging evidence for the
claims we have made.

\appendix
\section{The Heun equation\label{AppHeun}}

We used the convention a bit different from the NIST handbook \cite{NIST} to relate the equation to gauge theory quantities,
therefore in order to avoid confusion we go through some details on different forms of the Heun equation.

The Heun equation is the general second order Fuchsian linear differential equation with four regular singularities, it is an extension of the hypergeometric equation.
It can be written in several different forms, the most familiar form is
\be \f{d^2}{dz^2}w(z)+(\f{\gamma}{z}+\f{\eta}{z-1}+\f{\lambda}{z-q})\f{d}{dz}w(z)+\f{\alpha\beta z-Q}{z(z-1)(z-q)}w(z)=0,\ee
with $\alpha+\beta+1=\gamma+\eta+\lambda$, $q$ is called the singularity parameter, $\alpha,\beta,\gamma,\eta,\lambda$ are called the exponent parameters, and $Q$ is called the accessory parameter. There are four regular singularities at $0,q,1,\infty$, the merging of regular singularities gives the Confluent Heun equations with irregular singularities.

Define a new function $\widetilde{w}(z)$by
\be w(z)=z^{-\gamma/2}(z-1)^{-\eta/2}(z-q)^{-\lambda/2}\widetilde{w}(z),\ee then we can write the Heun equation in the normal form,
\be  \f{d^2}{dz^2}\widetilde{w}(z)-[\f{D}{z^2}+\f{F}{(z-q)^2}+\f{E}{(z-1)^2}+\f{(1-q)B-qA}{z(z-1)}+\f{q(1-q)(A+B)}{z(z-1)(z-q)}]\widetilde{w}(z)=0.\ee
with $A+B+C=0$. The parameters $A,B,C,D,E,F$ are related to $\alpha, \beta, \gamma, \eta,\lambda$ by
\beq &\quad&A=\f{2Q-q\gamma\eta-\gamma\lambda}{2q},\quad B=\f{2Q-2\alpha\beta+\gamma\eta+\lambda\eta-q\gamma\eta}{2(1-q)},\nn\\
&\quad& D=\f{1}{2}\gamma(\f{1}{2}\gamma-1),\qquad E=\f{1}{2}\eta(\f{1}{2}\eta-1),\qquad F=\f{1}{2}\lambda(\f{1}{2}\lambda-1).\eeq
Compare with the equation (\ref{equationVep}) obtained from CFT/gauge theory, we can identify the parameters $\gamma,\eta,\lambda, \alpha,\beta,Q$
with $m_0,m_1,\mt_0,\mt_1,q,\ep$ of the SU(2) $N_f=4$ theory. The identification is not unique, we have
\beq  &\gamma&=\f{2\mt_1}{\ep}+1,\qquad \m{or}\qquad -\f{2\mt_1}{\ep}+1,\nn\\
          &\eta&=\f{2m_0}{\ep},\qquad \m{or}\qquad -\f{2m_0}{\ep}+2,\nn\\
         &\lambda&=\f{2m_1}{\ep},\qquad \m{or}\qquad -\f{2m_1}{\ep}+2,\label{mass2coupling1}\eeq
and two more relations,
\begin{flalign}
&\alpha\beta=\f{\gamma\eta+\gamma\lambda+\eta\lambda}{2}+\f{m_0}{\ep}(\f{m_0}{\ep}-1)+\f{m_1}{\ep}(\f{m_1}{\ep}-1)-\f{\mt_0^2}{\ep^2}+\f{\mt_1^2}{\ep^2},&\nn\\
&Q=-(1-q)\f{\Theta}{\ep^2}+\f{q\gamma\eta+\gamma\lambda}{2}+q[\f{m_0}{\ep}(\f{m_0}{\ep}-1)+\f{m_1}{\ep}(\f{m_1}{\ep}-1)-\f{\mt_0^2}{\ep^2}+\f{\mt_1^2}{\ep^2}].&\label{Q2Theta}\end{flalign}
Terms involving masses in $\alpha\beta$ and $Q$ do not depend on parameter identification choice of (\ref{mass2coupling1}).

We can transform the equation to the elliptic form for the study of the TV potential. Define the coordinate
$\zeta$ by \be z=q\times\sn^2(\zeta,q),\label{z2zeta}\ee and define a new function $\widehat{w}(\zeta)$ by
\be w(z)=(\sn\zeta)^{(1-2\gamma)/2}(\dn\zeta)^{(1-2\eta)/2}(\cn\zeta)^{(1-2\lambda)/2}\widehat{w}(\zeta),\ee
where the Jacobi elliptic functions $\sn(\zeta,q)$ etc. depend on the nome $q$.
The quarter periods of $\sn(\zeta,q)$ etc. are complete elliptic integral $K(q), iK^{'}(q)$ (the same function as in (\ref{completelliptic}), but with different arguments due to mass deformation). Then the Heun equation transforms to the form
\be \f{d^2}{d\zeta^2}\widehat{w}(\zeta)+\Bigg(h-b_0q\sn^2\zeta-b_1q\f{\cn^2\zeta}{\dn^2\zeta}-b_2\f{1}{\sn^2\zeta}-b_3\f{\dn^2\zeta}{\cn^2\zeta}\Bigg)\widehat{w}(\zeta)=0,\label{HeunDarboux}\ee
where \beq h&=&-4Q+(\gamma+\lambda-1)^2+q(\gamma+\eta-1)^2,\quad b_0=-4\alpha\beta+(\gamma+\eta+\lambda-\f{1}{2})(\gamma+\eta+\lambda-\f{3}{2}),\nn\\
b_1&=&(\eta-\f{1}{2})(\eta-\f{3}{2}),\qquad
b_2=(\gamma-\f{1}{2})(\gamma-\f{3}{2}),\qquad
b_3=(\lambda-\f{1}{2})(\lambda-\f{3}{2}).\label{h2Q}
\eeq
This is the equation appears in Darboux's work \cite{darboux}. In the limit $q\to0$ and other quantities remain finite,
the potential reduces to the P\"{o}schl-Teller  potential.

In order to further transform the equation to the Weierstrass
elliptic form, we define variables $x$ by \be
\f{\zeta+iK^{'}}{(e_1-e_2)^{1/2}}=x, \quad\m{i.e.}\quad
\sn^2(\zeta,q)=\f{\wp(x,p)-e_2(p)}{e_3(p)-e_2(p)}.\ee
The half periods of $\wp(x)$ are $\omega_1,\omega_2$, we have $K=(e_1-e_2)^{1/2}\omega_1, iK^{'}=(e_1-e_2)^{1/2}\omega_2$.
The nome $p=\exp(2\pi i\f{\omega_2}{\omega_1})=\exp(-2\pi\f{K^{'}}{K})$ is related to $q$ as in (\ref{qprelation}).
Then we transform the equation (\ref{HeunDarboux}) to the following form, \be
\f{d^2}{dx^2}\widehat{w}(x)+\Bigg(E-\sum_{i=0}^{3}b_i\wp(x+\omega_i))\Bigg)\widehat{w}(x)=0,\label{HeunW}\ee
where  \beq E&=&(e_1-e_2)h+e_2\sum_ib_i\nn\\
&=&4(e_2-e_1)Q-4e_2\alpha\beta+e_1(\gamma+\lambda-1)^2+e_2(\eta+\lambda-1)^2+e_3(\gamma+\eta-1)^2.\label{H2Q}\eeq

The stationary points of the potential are determined by the condition $\sum_ib_i\p_x\wp(x+\omega_i)=0$,
if rewritten in the variable $z$ it is a polynomial equation of degree six $Q_6(z)=0$.
The six solutions are in correspondence to the six solutions for the discriminant equation $\mathcal{D}(P_4)(\Xi)=0$, discussed in the next two Appendixes.

\section{Iterative solution for $P_4(z)=0$\label{AppP4}}

{\bf The degenerate points of $P_4(z)$}

Let us start from some basic facts about quartic polynomial. For a quartic polynomial defined in the complex domain
\be P_4(z)=az^4+bz^3+cz^2+dz+e,\ee
we can associated an elliptic curve to the polynomial, $y^2=P_4(z)$.
The shape of the curve is controlled by the modulus parameter (\ref{modulus}), which is given by the cross ratio of the roots of the equation $P_4(z)=0$.

These elliptic curves are used in the Seiberg-Witten theory to determine the dynamics of the gauge theory,
when the curve degenerates the gauge theory has a weak coupling description.
It happens when two of the roots collide, making either the modulus $k^2$ or its complementary modulus $k^{'2}$ small.
$k^2$ and $k^{'2}$ are related to the gauge coupling and the dual gauge coupling, respectively.
In this paper we need to deal with an integral that is similar to the integration of the Seiberg-Witten form,
the spirit is the same: when the curve degenerate the integral can be written as an asymptotic expansion.
The condition for the degeneration is given by the vanishing of the discriminant of the polynomial,
\beq \mathcal{D}(P_4)&=&b^2c^2d^2-4ac^3d^2-4b^3d^3+18abcd^3-27a^2d^4-4b^2c^3e+16ac^4e\nn\\
&\quad&+18b^3cde-80abc^2de-6ab^2d^2e+144a^2cd^2e-27b^4e^2+144ab^2ce^2\nn\\
&\quad&-128a^2c^2e^2-192a^2bde^2+256a^3e^3. \eeq

Now we specific to the quartic polynomial used in our story, it is
(see \cite{mt1006, em1006})
\beq P_4(z)&=&\mt_0^2z^4+(-\Xi+q\Xi-2q\mt_0^2+2qm_1^2+m_0^2-\mt_0^2-m_1^2-\mt_1^2)z^3\nn\\
&\quad&+(\Xi-q^2\Xi+q^2\mt_0^2-q^2m_1^2-2qm_0^2+2q\mt_0^2-2qm_1^2+2q\mt_1^2+m_1^2+\mt_1^2)z^2\nn\\
&\quad&+(-q\Xi+q^2\Xi+q^2m_0^2-q^2\mt_0^2+q^2m_1^2-q^2\mt_1^2-2q\mt_1^2)z+q^2\mt_1^2.\label{poly4}\eeq
The polynomial does not define the Seiberg-Witten curve of the super QCD, however,
the spirit of finding the degenerate points of the associated curve to carry out the asymptotic expansion is similar.
The equation $P_4(z)=0$ has general algebraic solutions, but their closed form are too cumbersome to analyse in practice,
however, here we can turn some parameters very small and asymptotic solutions can be found.
Starting from a weak coupling UV theory, we have $q$ always small.
We can also turn the scalar v.e.v which controls the magnitude of $\Xi$, therefore we can make $\Delta=\Xi-\Xi_*$ either very large or very small,
where $\Xi_*$ is a degeneration point of the polynomial $P_4(z)$.

The asymptotic solution can be expanded by small parameters, the corresponding expansion coefficients can be solve by an iterative method,
if we correctly choose the leading order solution and the subsequent expansion pattern. The iterative method works as a self-checking process,
it only works well when we choose the right expansion pattern, otherwise it always ends in an obstruction.

The location of the degeneration point $\Xi_*$ where
$\mathcal{D}(P_4)(\Xi_*)=0$, determines the nature of the asymptotic
solution, therefore we need to find its value and do asymptotic
expansion near this point. For generic masses, the discriminant of
$P_4(z)$ is a large polynomial of degree six for $\Xi$. Therefore
there are six singularities in the $\Xi$ space at finite position,
each of them is of weight one. Their locations are
\beq \Xi_*&\sim&-m_1^2-\mt_1^2+(m_0\pm\mt_0)^2+\mathcal{O}(m^2q),\nn\\
&\quad& \pm 2m_1\mt_1+\mathcal{O}(m^2q),\nn\\ &\quad&-m_1^2-\mt_1^2\pm2(m_0^2-\mt_0^2)^{1/2}(m_1^2-\mt_1^2)^{1/2}q^{1/2}+\mathcal{O}(m^2q).\label{5singu}\eeq
Then the shifted coordinates defined by,
\be \widetilde{\Xi}=m_1^2+\mt_1^2+\f{2q}{1-q}m_0m_1+\Xi,\ee
has singularities at
\beq \widetilde{\Xi}_*&\sim&(m_0\pm\mt_0)^2+\mathcal{O}(m^2q),\nn\\
&\quad&(m_1\pm\mt_1)^2+\mathcal{O}(m^2q),\nn\\ &\quad&\pm2(m_0^2-\mt_0^2)^{1/2}(m_1^2-\mt_1^2)^{1/2}q^{1/2}+\mathcal{O}(m^2q).\eeq
Note that from (\ref{theta2prepotential}) we have $\widetilde{\Xi}_*\sim q\p_q\mathcal{F}(q)$.

Note that $(m_0\pm\mt_0), (m_1\pm\mt_1)$ are the bare masses
$\mu_1,\mu_2,\mu_3,\mu_4$, hence we conclude the first four
singularities, denoted as $\Xi_{*1,2,3,4}$, are associated with the
flavor $\mu_i$ which becomes massless at the singularity $\Xi_{*i}$
of the Coulomb moduli, they are in the semiclassical region in the
moduli space. The asymptotic expansion in this region is formula
(\ref{XIallorder}), when $a\sim\mu_i$ then $\Xi$ is at the
corresponding singularity. The singularity $\Xi_{*i}$ would be
pushed to infinity under the decoupling limits involving the flavor
$\mu_i$, and remain finite for other mass decoupling limits.

In the semiclassical region, $\Xi$ is large, we should have set $\Xi=\Xi_{*1,2,3,4}+\Delta$ with $\Delta$ large,
but in section \ref{TVspectrum} we treat $\Xi$ itself as a large quantity. We can divide all the coefficients of the polynomial (\ref{poly4}) by $\Xi$, and the degeneration at $\Xi\sim\infty$ is obvious.

Other two singularities in (\ref{5singu}) are associated with magnetic and dyonic particles, in the mass decoupling limit they survive as the corresponding strong coupling singularities for the gauge theory with less flavors.

Before we proceed to solve, it is helpful to notice the following properties the solutions should satisfy for generic cases of parameters:
\begin{itemize}
\item[(1.)] When $\mt_0\to0$, the equation reduces to third order, it means a root is pushed to infinity.
\item[(2.)] When $\mt_1\to0$, the equation has a root $z=0$.
\item[(3.)] When $m_0\to0$, the equation has a root $z=1$
\item[(4.)] When $m_1\to0$, the equation has a root $z=q$.
\end{itemize}

{\bf Large $\Xi$ solution}

For large $\Xi$, the leading order of the polynomial reduces to a cubic one, $P_4(z)=\Xi(q-1)z(z-q)(z-1)+\cdots$, therefore the leading order solution is given by the equation $z(z-q)(z-1)=0$ which gives three roots at $0,q,1$, the forth root $z_4$ is near infinity. We use small quantity $\f{1}{\Xi}$ to control the expansion,
with a little of guess, we have the large expansion of all four roots as presented in (\ref{largeexp}).
The subleading coefficients are iteratively solved. Take $z_2$ as the example, if we already got the coefficient $f_2^j, 2\le j\le n$, substitute the trial solution
\be z_2=q+\sum_{j=1}^{n}\f{f_2^j}{\Xi^j}+\f{f_2^{n+1}}{\Xi^{n+1}},\ee
into the polynomial $P_4(z)$,  then the leading order non-zero coefficient in the $\f{1}{\Xi}$ expansion of $P_4(z)$ is a function of the form $c_1+c_2f_2^{n+1}$ with $c_{1,2}$ functions of masses and $q$. We can solve the equation at this order by setting $c_1+c_2f_2^{n+1}=0$ therefore obtain the coefficient $f_2^{n+1}$.
In this way we can iteratively solve all $f_i^j$.
We give the first few of them.
\beq
f_1^2&=&\f{q\mt_1^2(qm_0^2+qm_1^2-q\mt_0^2-\mt_1^2)}{(1-q)^2},\nn\\
f_1^3&=&-(1-q)^{-3}q\mt_1^2[q^2(m_0^4+2m_0^2m_1^2+m_1^4-2m_0^2\mt_0^2-2m_1^2\mt_0^2+\mt_0^4+m_0^2\mt_1^2)\nn\\
&\quad&-3q\mt_1^2(m_0^2+m_1^2-\mt_0^2)+\mt_1^2(m_1^2+\mt_1^2)],\nn\\
f_2^2&=&\f{qm_1^4(1-2q)}{1-q},\nn\\
f_2^3&=&-\f{qm_1^4[q^2(4m_1^2+\mt_0^2)+q(m_0^2-4m_1^2-\mt_0^2-\mt_1^2)+m_1^2+\mt_1^2]}{(1-q)^2},\nn\\
f_3^2&=&\f{m_0^2[q(m_1^2-\mt_0^2+\mt_1^2)+m_0^2-m_1^2+\mt_0^2-\mt_1^2]}{1-q},\nn\\
f_4^0&=&\f{q\mt_0^2-2qm_1^2-m_0^2+m_1^2+\mt_1^2}{\mt_0^2},\nn\\
f_4^1&=&-\f{q^2m_1^2-qm_0^2-qm_1^2+q\mt_1^2+m_0^2}{1-q}.
\eeq
Note that $q$ is automatically incorporated into the expansion in a reasonable form. These solution satisfy the properties of limits (1.)-(4.).

{\bf Small $\Delta$ solution}

Around the magnetic point, we set $\Xi=\Xi_{mag}+\Delta$, substitute this into $P_4(z)$ we get a polynomial double expanded w.r.t. to $q$ and $\Delta$.
In this polynomial the $q$-expansion is infinite order, therefore the iterative solution would be more involved than the case of large $\Xi$.
We get the following leading order solution, and the general expansion pattern,
\begin{flalign}
&z_1=-\f{m_0^2-\mt_0^2}{\mt_0^2}+\f{2m_0^2(m_1^2-\mt_1^2)^{1/2}}{\mt_0^2(m_0^2-\mt_0^2)^{1/2}}q^{1/2}+\sum_{n=2}^{\infty}c_*q^{n/2}&\nn\\
&\qquad+\left(\f{m_0^2}{\mt_0^2(m_0^2-\mt_0^2)}+\f{4m_0^2(m_1^2-\mt_1^2)^{1/2}}{(m_0^2-\mt_0^2)^{5/2}}q^{1/2}+\sum_{n=2}^{\infty}c_*q^{n/2}\right)\Delta &\nn\\
&\qquad+\sum_{m=2}^{\infty}(\sum_{n=0}^{\infty}c_*q^{n/2})\Delta^m,&\nn\end{flalign}
\begin{flalign} &z_2=-\f{(m_1^2-\mt_1^2)^{1/2}}{(m_0^2-\mt_0^2)^{1/2}}q^{1/2}+\f{m_1^2(m_0^2-\mt_0^2)^2-m_0^2(m_1^2-\mt_1^2)^2}{(m_0^2-\mt_0^2)^2(m_1^2-\mt_1^2)}q+\sum_{n=3}^{\infty}c_*q^{n/2}&\nn\\
&\qquad+\left(-\f{(m_1^2-\mt_1^2)^{1/4}}{(m_0^2-\mt_0^2)^{3/4}}q^{1/4}-\f{2m_0^2(m_1^2-\mt_1^2)^{3/4}}{(m_0^2-\mt_0^2)^{9/4}}q^{3/4}+\sum_{n=2}^{\infty}c_*q^{n/2+1/4}\right)\Delta^{1/2}&\nn\\
&\qquad+\sum_{m=1}^{\infty}(\sum_{n=0}^{\infty}c_*q^{n/2})\Delta^{m}+\sum_{m=1}^{\infty}(\sum_{n=0}^{\infty}c_*q^{1/4-m/2+n/2})\Delta^{m+1/2},&\nn\end{flalign}
\begin{flalign} &z_3=-\f{(m_1^2-\mt_1^2)^{1/2}}{(m_0^2-\mt_0^2)^{1/2}}q^{1/2}+\f{m_1^2(m_0^2-\mt_0^2)^2-m_0^2(m_1^2-\mt_1^2)^2}{(m_0^2-\mt_0^2)^2(m_1^2-\mt_1^2)}q+\sum_{n=3}^{\infty}c_*q^{n/2}&\nn\\
&\qquad-\left(-\f{(m_1^2-\mt_1^2)^{1/4}}{(m_0^2-\mt_0^2)^{3/4}}q^{1/4}-\f{2m_0^2(m_1^2-\mt_1^2)^{3/4}}{(m_0^2-\mt_0^2)^{9/4}}q^{3/4}+\sum_{n=2}^{\infty}c_*q^{n/2+1/4}\right)\Delta^{1/2}&\nn\\
&\qquad+\sum_{m=1}^{\infty}(\sum_{n=0}^{\infty}c_*q^{n/2})\Delta^{m}-\sum_{m=1}^{\infty}(\sum_{n=0}^{\infty}c_*q^{1/4-m/2+n/2})\Delta^{m+1/2},&\nn\end{flalign}
\begin{flalign}
&z_4=-\f{\mt_1^2}{m_1^2-\mt_1^2}q-\f{2m_1^2\mt_1^2(m_0^2-\mt_0^2)^{1/2}}{(m_1^2-\mt_1^2)^{5/2}}q^{3/2}+\sum_{n=4}^{\infty}c_*q^{n/2}&\nn\\
&\qquad-\left(\f{m_1^2\mt_1^2}{(m_1^2-\mt_1^2)^3}q+\f{4m_1^2\mt_1^2(m_0^2-\mt_0^2)^{1/2}(m_1^2+\mt_1^2)}{(m_1^2-\mt_1^2)^{9/2}}q^{3/2}+\sum_{n=4}^{\infty}c_*q^{n/2}\right)\Delta &\nn\\
&\qquad+\sum_{m=2}^{\infty}(\sum_{n=1}^{\infty}c_*q^{n})\Delta^{m}, &\end{flalign}
where we use $c_*$ to denote any coefficients which are
functions of only mass parameters. These solution also satisfy the
properties of limits (1.)-(4.). In principle the roots can be solved
order by order, but the coefficients become increasingly lengthy.

\section{Stationary points of Treibich-Verdier potential\label{Appstationary}}

In this Appendix we show a fact that the stationary points of the TV
potential are in one-to-one correspondence with the degenerate
points of the polynomial $P_4(z)$. From the equation of Darboux form
(\ref{HeunDarboux}), we obtain the stationary condition for the potential $\p_\zeta V(b_i,\zeta)=0$,
then using the relation of $\zeta$ and $z$ (\ref{z2zeta}) we get the condition $Q_6(z)=0$ which is convenient for iterative solving.
Let us work with the unshifted parameters that appear in (\ref{equationV0}), we have
\beq Q_6(z)&=&\mt_0^2z^6-2(1+q)\mt_0^2z^5-[m_0^2-\mt_0^2-q(m_0^2+4\mt_0^2+m_1^2-\mt_1^2)-q^2(\mt_0^2-m_1^2)]z^4\nn\\
&\quad&+2q[(m_0^2-\mt_0^2-m_1^2+\mt_1^2)-q(m_0^2+\mt_0^2-m_1^2-\mt_1^2)]z^3\nn\\
&\quad&+q[m_1^2-\mt_1^2-q(m_0^2-\mt_0^2+m_1^2+4\mt_1^2)+q^2(m_0^2-\mt_1^2)]z^2+2q^2(1+q)\mt_1^2z\nn\\
&\quad&-q^3\mt_1^2.\eeq
Applying the iterative method, we can solve the equation $Q_6(z)=0$ perturbatively.
Expanded by $q$, they appears in pairs,
\beq z_{*1,2}&=&\f{\mt_0\pm m_0}{\mt_0}\mp\f{m_0[(m_0\pm\mt_0)^2+m_1^2-\mt_1^2]}{2\mt_0(m_0\pm\mt_0)^2}q+\mathcal{O}(q^2),\nn\\
z_{*3,4}&=&\f{\mt_1}{\mt_1\pm m_1}q\pm\f{m_1\mt_1[m_0^2-\mt_0^2+(\mt_1\pm m_1)^2]}{2(m_1-\mt_1)^4}q^2+\mathcal{O}(q^3),\nn\\
z_{*5,6}&=&\pm\f{(m_1^2-\mt_1^2)^{1/2}}{(m_0^2-\mt_0^2)^{1/2}}q^{1/2}+\f{m_1^2(m_0^2-\mt_0^2)^2-m_0^2(m_1^2-\mt_1^2)^2}{(m_0^2-\mt_0^2)^2(m_1^2-\mt_1^2)}q\pm\mathcal{O}(q^{3/2}).
\eeq

For a stationary particle at the point $z_{*i}$, as it loses the kinetic energy, its energy is the potential energy,
\be h_{*i}=b_0q\sn^2\zeta+b_1q\f{\cn^2\zeta}{\dn^2\zeta}+b_2\f{1}{\sn^2\zeta}+b_3\f{\dn^2\zeta}{\cn^2\zeta}\Big|_{\zeta=\zeta(z_{*i})}.\ee
Then from the relation (\ref{Q2Theta})(\ref{h2Q}) we can compute the energy $\Theta_{*i}$(or $\Xi_{*i}$) at the stationary points $z_{*i}$.
The six $\Xi_{*1,2,3,4,5,6}$ associated to $z_{*1,2,3,4,5,6}$ appear in pairs too,
\beq \Xi_{*1,2}&=&-m_1^2-\mt_1^2+(m_0\pm\mt_0)^2+\f{m_0[(m_0\pm\mt_0)^2+m_1^2-\mt_1^2]}{m_0\pm\mt_0}q+\mathcal{O}(q^2),\nn\\
\Xi_{*3,4}&=&\pm2m_1\mt_1+\f{m_1[m_0^2-\mt_0^2+(m_1\pm\mt_1)^2]}{m_1\pm\mt_1}q+\mathcal{O}(q^2),\nn\\
\Xi_{*5,6}&=&-m_1^2-\mt_1^2\pm2(m_0^2-\mt_0^2)^{1/2}(m_1^2-\mt_1^2)^{1/2}q^{1/2}+\mathcal{O}(m^2q).\eeq
They are the same values as in (\ref{5singu}) obtained from solving
$\mathcal{D}(P_4)(\Xi)=0$, and they satisfy the relation
$P_4(\Xi_{*i}, z_{*i})=0$. Recall that the solutions of $P_4(\Xi,
z)=0$ are $z_{1,2,3,4}(\Xi)$, they depend on $\Xi$, when
$\Xi=\Xi_{*i}$ two roots among $z_{1,2,3,4}(\Xi)$ collide. There are
$6=C_4^2$ ways to collide two roots, exactly to a value $z_{*i}$.
Therefore the six stationary points $z_{*i}$ of the TV potential are
in one-to-one correspondence with the six weak coupling asymptotic
expansion points for gauge theory. So the situation is similar to
the Lam\'{e} potential we have studied \cite{wh1108}.

\section*{Acknowledgments}

I would like to thank  Andrei Mikhailov for useful discussions and comments on the draft,
Oscar Chacaltana, Niclas Wyllard for discussions. This work is supported by the FAPESP No. 11/21812-8, through IFT-UNESP.

\end{document}